\documentclass[12pt]{article}
\usepackage{amssymb}
\usepackage{amsmath}
\usepackage{latexsym}
\usepackage{epsfig}
\usepackage{subfigure} 
\usepackage[font=footnotesize,labelfont=bf]{caption}

\newif{\ifcomentarios}
\comentariosfalse

\begin{document}

\author{Vitor H. Sanches, Dhyan V. H. Kuraoka, Pedro R. de Almeida, Carla
Goldman \footnote{Email: carla@if.usp.br. Instituto de F\'{\i}sica - Universidade de S\~{a}o Paulo, CEP 05508-090, S%
\~{a}o Paulo-SP, Brasil.}}
\title{A phenomenological analysis of eco-evolutionary coupling under dilution.}
\date{May 11, 2017 }
\maketitle

\begin{abstract}
  \noindent
Evolutionary dynamics experienced by mixed microbial populations of
cooperators and cheaters has been examined in experiments in the literature
using a protocol of periodic dilution to investigate the properties of
resilience and adaptability to environmental changes. Data depicted on an
appropriate phase diagram indicate, among other features, a stable
equilibrium point at which cooperators and cheaters coexist [A. Sanchez, J.
Gore, PLOs Biology, 11 (4), e1001547 (2013)]. We present here a
phenomenological analysis of these data focusing on an eco-evolutionary-game
perspective. To that end, we work on an extension of the model proposed by
Tao and Cressman Y. Tao, R. Cressman, Bull. Math. Biol. 69, 1377 - 1399
(2007). It\'{}
s original version takes into account changes of the total population
density while the individuals experience pairwise Prisoner\'{}s Dilemma game. The extension devised here contains a dilution factor to be
conform with the experimental procedure, in addition of a term accounting
for Allee effects. Differently from other descriptions proposed in similar
contexts, however, the model here does not account for assortative
encounters, group or kin selection. Nonetheless, it describes surprisingly
well both qualitatively and quantitatively the features of the observed
phase diagram. We discuss these results in terms of the behavior of an
effective payoff matrix defined accordingly. 
\end{abstract}

\section{Introduction}

Ecological and evolutionary dynamics taking place at similar timescale may
drive certain interacting populations of individuals to an eco-evolutionary
feedback, a situation under which ecological limitations may control
evolutionary changes and vice-versa \cite{Post Palkovacs 2009}. Microbes
have always been considered as promising candidates to exhibit
eco-evolutionary cyclic process although the experimental confirmation of
such expectations occurred only a few years ago investigating mixed
population of budding yeast \textit{Saccharomyces cerevisiae }\cite{Sanchez
and Gore PLOs 2013}. The individuals in these populations that posses the
gene SUC2\textit{\ cooperate} since they codify for the production of
\textit{invertase, }an important enzyme to the process of the hydrolysis of
sucrose into fructose and glucose, helping yeast to improve the use of the
sucrose present in the medium\textit{. }The other individuals considered in
the experiments, the "\textit{cheaters",} do not have the gene SUC2. Despite
of this, cheaters may still be benefited by the products, sucrose and
fructose left available in the medium by the cooperators, as a \textit{%
public good}, avoiding in this way the cost of production. Data released
from these experiments show the behavior of the fraction (frequency) $p(t)$
of cooperators present in the sample at each time $t$, analyzed as a
function of the total population density $N(t)$. In special, one can
identify in these data a stable \textit{coexistence equilibrium} reached by
the two populations in the long-time regime. These results drew considerable
attention in the literature due to the their relevance in the context of the
\textit{dilemma of cooperation} in biology \cite{Allen Nowak PLOs 2013}.

In another set of related experiments \cite{Dai and Gore Science 2012} one
examined questions regarding changes on resilience of single populations
consisting only of the referred cooperators subject to dilution, the
strength of which has been used as a controlling parameter. According to the
authors, this procedure shall be equivalent to produce changes on the
intrinsic mortality rate of the population since it consists on a periodic
removal of a certain number of individuals in proportion to the quantity
present at each instant of time. The evolution of the single population of
cooperators was examined there under these conditions for diverse initial
population sizes and different values of a defined dilution factor $\theta $
which also determines the behavior of the observed equilibrium points. A
turning point bifurcation is identified in these data, occurring at the
value of $\theta $ for which these equilibrium points coalesce \cite{Dai and
Gore Science 2012}.

Traditionally, such questions regarding the evolution of cooperation in
competing populations have being addressed theoretically based on \textit{%
evolutionary game theory} \cite{nowak sigmund 2004} coupled to classical
models from ecology (Lotka-Volterra) to take into account variations of the
size $N$ of the entire population under environment constraints, see for
example \cite{Hauert Holmes Doebeli}, \cite{Feng Zhang and Cang Hui}, \cite%
{Frey}. This kind of formulation extends the dynamics described by the
\textit{replicator equation} conceived on the basis of evolutionary game
theory for \textit{pairwise encounters}, to describe exclusively
evolutionary aspects of populations of constant sizes \cite{Nowak livro}.

In general terms, the long-time behavior observed in the experimental data
in \cite{Sanchez and Gore PLOs 2013}\textbf{\ }has been predicted by some of
these \textit{eco-evolutionary models} \cite{Hauert Holmes Doebeli}, \cite%
{Feng Zhang and Cang Hui}. Yet, coexistence of cooperators and cheaters has
been achieved using the \textit{Prisoner%
\'{}%
s Dilemma Game} (PDG) in such context only under favorable conditions for
assortative encounters \cite{Bergstrom 2003} or cooperation among
individuals in groups of all possible sizes.

Conversely, the eco-evolutionary model proposed in Ref.\cite{Requejo Camacho
PRL 2012} has been conceived in the absence of structured populations, kin
selection, or assortative encounters. Yet, limitations of resources are not
described by Lotka-Volterra equations, but are introduced directly into the
elements of the payoff matrix that defines the game, expressing rewards and
costs in terms of resources exchanged between each pair of individuals. The
dynamic payoffs constructed in this way set conditions under which the
population composed of two defined types of individuals is driven to a
stable coexistence. The nature of both the competition investigated by the
model and the predicted equilibrium, however, show no correspondence with
the system investigated in the experiments mentioned above.

A rather simple model has been proposed by the same authors of the
experiments to explain their data. It is based entirely on Lotka-Volterra
equations expressing competition for external resources but in the absence
of a game. Because there are indications in the data suggesting that the
maximum per-capita growth rate (or \textit{intrinsic growth rate}) differs
for each population, such differences are introduced into their model as the
only way to distinguish cooperators from cheaters. In addition, the authors
assume that these intrinsic growth rates change when population reaches a
certain critical density $N=N_{C}$, introduced as an external input. Thus,
the complete model devised there comprises four equations to describe the
time variation of the densities of cooperators and cheaters, two of them for
total population density values $N<$ $N_{C}$ and the other two with
different parameters, for $N>$ $N_{C}$. Dilution is not explicitly
introduced into these equations but it is implemented in the numerical
simulation for studying the dynamics.

Here, we resume the more traditional eco-evolutionary view on the basis of
the model introduced by Tao and Cressman (TC) \cite{Tao Cressman 2007}. This
model has been formulated originally to study stochastic effects on the
evolution of competitive populations. Its deterministic limit accounts for
variations of the total density $N$ through Lotka-Volterra dynamics while
individuals experience pairwise PDG in the absence of assortative
encounters, group or kin selection. The long time dynamics of this model
does not predict stability of cooperators, as expected.\textbf{\ }In the
present work, we extend the deterministic version of this model to proceed
into a phenomenological analysis of the aforementioned experiments. We
explicitly introduce into the original TC model an extra factor to account
for the dilution protocol. Also in addition to a Lotka-Volterra contribution
that enters in the original formulation as a $N$-dependent background
fitness, we take into account Allee effects \cite{Allee} and differences on
the intrinsic growth rates of each population which, however, are conserved
along the entire dynamics.

In Section 2 we explain our model and argue that it meets the requirements
to describe specificities of the experiments. The analysis presented in
Section 3 indicate that both its equilibrium and dynamic properties
reproduce remarkably well the details reported in both experimental works
\cite{Sanchez and Gore PLOs 2013} and \cite{Dai and Gore Science 2012}. The
consistence between these two phenomenologycal studies suggests that the
model is surprisingly robust. To our knowledge, this is the first time that
eco-evolutionary feedback resulting in coexistence is predicted on the basis
of the original pairwise PDG. The analysis presented in Section 4 in terms
of the properties of an effective game defined accordingly, allows us to
understand the maintenance of cooperation in the long time regime in terms
of a Nash equilibrium between two competing populations that depends on $N$.
Concluding remarks are in Section 5.

\section{Eco-evolutionary model with dilution}

We consider the time evolution of well mixed microbial populations of
interacting cooperators and defectors (cheaters). Let $n_{i}(t)$ be the
density (number of individuals per unit volume) of cooperators $(i=1)$ and
cheaters $(i=2)$ present in the mixture of fixed volume at each time $t$.
Then $N(t)=n_{1}(t)+n_{2}(t)$ is the total population density and $p_{i}(t)=$%
\emph{\ }$n_{i}(t)/N(t),$ $i=1,2$ \ the corresponding frequencies\emph{. }We
express the time variation of individual densities as

\begin{equation}
\begin{array}{l}
\overset{\cdot }{n}_{1}=n_{1}\left\{\left(\frac{N}{A}%
-1\right)[(a_{11}p_{1}+a_{12}p_{2}+\lambda (1-\beta N)]-\nu \right\} \\
\\
\overset{\cdot }{n}_{2}=n_{2}\left\{\left(\frac{N}{A}%
-1\right)[(a_{21}p_{1}+a_{22}p_{2})+\delta (1-\beta N)]-\nu \right\}%
\end{array}
\label{modelo 6}
\end{equation}%
The curly brackets include the total \textit{fitness} of each population.
The dependence of the fitness on $\beta $ and on the parameters $a_{ij},$ $%
i,j=1,2$ comprise precisely the original model proposed in Ref. \cite{Tao
Cressman 2007}.\textbf{\ }$\beta $ can be interpreted as the inverse of
\textit{carrying capacity} of the system. Thus, the factor $(1-\beta N)$ in
each of the equations above represents the usual Lotka-Volterra factor
imposing limitations on population growth at relative high values of $N$,
due to environment constraints. As noticed by the authors \cite{Tao Cressman
2007} it can be ascribed, in the context, to a \textit{background fitness}
for both populations. The \textit{payoff matrix} elements $a_{ij}$ are
constants representing the effects on fitness due to the interactions
between any pair of individuals that play a \textit{game }chosen to coicide
with the \textit{Prisoner's Dilemma Game (PDG}). Each $a_{ij}$ is
interpreted as the reward for each individual, either a cooperator (Co) or a
cheater (Ch), that undergoes pairwise disputes. Following the usual
representation \cite{Nowak livro} the PDG payoff matrix is expressed as:

\begin{equation}
  \bordermatrix{~ & Co & Ch \cr
                    Co & B-C & -C \cr
                    Ch & B & 0 \cr}
\label{payoff matrix}
\end{equation}

with $B,C>0$ and $B>C.$ The constant $B$ is a measure of the benefit
received by a player in disputes (encounters) with a cooperator. In the
present case, it corresponds to the benefit after using the considered
substances (fructose and glucose). $C$ is a measure of the total cost
associated to the production of these substances which is spent by
cooperators only.

The remaining factors in (\ref{modelo 6}) modify the original equations in
Ref. \cite{Tao Cressman 2007}, as explained next:

\textit{i)}\textbf{\ }The factor $(\frac{N}{A}-1)$ was introduced following
the suggestions made by the authors of the experiments \cite{Sanchez and
Gore PLOs 2013}, based on evidences in the data indicating that a minimum
population size is needed to transform the environment into glucose rich.
Only under this favorable condition the benefit promoted by the cooperators
would be shared among all other individuals present in the population at
each time. This is introduced into the equations above as a strong \textit{%
Allee} like effect that depends on the parameter $A$ setting the scale at
which such effects are expected to be relevant \cite{Allee}. It affects
negatively the rates for small population i.e., far from the carrying
capacity of the system, $\beta ^{-1}$.

\textit{ii)} The two constants $\lambda $ and $\delta $ for cooperators and
cheaters, respectively distinguishes the \textit{intrinsic} per-capita
growth rates. Such difference has also been noticed in the experimental data
and thus included by the authors in the analysis accompanying the
experimental report \cite{Sanchez and Gore PLOs 2013}.

\textit{iii)} The factor linear in $n_{1}$ in the first equation (or $n_{2}$
in the second), with a proportionality constant $\nu $ intends to reproduce
the protocol of dilution introduced in the experiments \cite{Sanchez and
Gore PLOs 2013},\textbf{\ } \cite{Dai and Gore Science 2012} . According to
this, the quantity of individuals removed after each\ interval of 24 hours
corresponds to a fraction of the total present in the mixture at that
instant of time.

\emph{\ }Adding the two equations in (\ref{modelo 6}) with the definition $p$
$\equiv p_{1}$ and using the payoffs as in (\ref{payoff matrix}) it results
\begin{equation}
\overset{\cdot }{N}=N\left\{ \left( \frac{N}{A}-1\right) [p(B-C)+(1-\beta
N)(p(\lambda -\delta )+\delta )]-\nu \right\}  \label{dN/dt}
\end{equation}%
The expression for the time evolution of the frequency of cooperators $p$
can then be derived with the aid of the identity $\overset{\cdot }{n}_{1}=%
\overset{\cdot }{p}N+p\overset{\cdot }{N}$, yielding%
\begin{equation}
\overset{\cdot }{p}=p(1-p)\left( \frac{N}{A}-1\right) \left[ -C+(\lambda
-\delta )(1-\beta N)\right] .  \label{dp/dt}
\end{equation}

The system comprising the above non-linear equations (\ref{dN/dt}) and (\ref%
{dp/dt}) couples the two quantities $N$ and $p$ which are the focus of the
cited experiments. This pair of equations constitutes the basis for the
phenomenological analysis we perform hereafter. We emphasize that for
intrinsic growth rates $\lambda =\delta =1$ and in the absence of both,
dilution $(\nu =0)$ and Allee effects, these equations reduce to the
deterministic version of the equations proposed by Tao and Cressman \cite%
{Tao Cressman 2007}\textbf{.} These same conditions are necessary (but not
sufficient) for this model to reduce to the deterministic equations of Frey
and col. \cite{Frey}\ to examine the transient behavior of cooperators
\textbf{\footnote{%
The two sets of equations coincide if, in addition to the conditions stated
in the text, both the strength of selection $s$ and the global birth fitness
function $\ $introduced in Ref. \cite{Frey} are taken equal to the unity.}}.

\section{Analysis}

The non-trivial equilibrium points of the dynamics described by Equations (%
\ref{dN/dt}) and (\ref{dp/dt}), denoted as $(N_{(i)},p_{(i)}),$ $i=1,2,3$
are listed below:%
\begin{equation}
\begin{array}{lll}
p_{(1)}=0 & N_{(1)}^{\pm }=\tfrac{A}{2\beta }\left[ \left( \tfrac{1}{A}%
+\beta \right) \pm \sqrt{\left( \tfrac{1}{A}-\beta \right) ^{2}-4\tfrac{%
\beta \nu }{A\delta }}\right] & (a) \\
&  &  \\
p_{(2)}=1 & N_{(2)}^{\pm }=\tfrac{A}{2\beta }\left[ \left( \tfrac{1+\left(
B-C\right) /\lambda }{A}+\beta \right) \pm \sqrt{\left( \tfrac{1+\left(
B-C\right) /\lambda }{A}-\beta \right) ^{2}-4\tfrac{\beta \nu }{A\lambda }}%
\right] & (b) \\
&  &  \\
p_{(3)}=\tfrac{1}{B}\left( \tfrac{\nu }{\left( N_{(3)}/A\right) -1}-\tfrac{%
C\delta }{\lambda -\delta }\right) & N_{(3)}=\tfrac{1}{\beta }\left( 1-%
\tfrac{C}{\lambda -\delta }\right) & (c)%
\end{array}
\label{pontos fixos}
\end{equation}%
In addition, the dynamics may drive the entire population to extinction
since $N=0$ is also an equilibrium for all $p\in \lbrack 0,1].$ The nature
of each of these non-trivial points $(N_{(i)},p_{(i)})$ shall be determined,
as usual, by linear analysis of Eqs. (\ref{dN/dt}) and (\ref{dp/dt})
examining the properties of the corresponding eigenvalues of the Jacobian
matrix:
\begin{equation}
\left. J\right\vert _{(N_{(i)},p_{(i)})}=\left(
\begin{array}{cc}
\frac{\partial \overset{\cdot }{N}}{\partial N} & \frac{\partial \overset{%
\cdot }{N}}{\partial p} \\
\frac{\partial \overset{\cdot }{p}}{\partial N} & \frac{\partial \overset{%
\cdot }{p}}{\partial p}%
\end{array}%
\right) _{(N_{(i)},p_{(i)})}  \label{jacobiano np}
\end{equation}%
evaluated at each of these points $(N_{(i)},p_{(i)})$. If not otherwise
specified, $\nu $ shall be taken as a controlling parameter in the analysis
of the two equilibrium points at the line $p=p_{(1)}=1$.

The numerical values for the remaining parameters are adjusted
phenomenologically by identifying each one of the equilibrium points
observed in \cite{Sanchez and Gore PLOs 2013}, namely $\{(5.96\times
10^{4},1),$\ $(516,1),(5.78\times 10^{4},0.086)\}$ with the corresponding
expressions in the set $\{(N_{(2)}^{+},1),$\ $%
(N_{(2)}^{-},1),(N_{(3)},p_{(3)})\}$ (\ref{pontos fixos}). The two points
located at $p=1$ indicate two monomorphic equilibria, and the one internal
characterize coexistence of the two populations. This procedure set the
values of the parameters that will be used in the subsequent analysis. The
results are compiled in Table 1.%
\begin{table}
  \center
  \caption{Parameters used in the simulation}
\begin{tabular}{cc}
Parameter & \text{Numerical Value} \\
B & 5.0799 \\
C & 1.00 \\
$\lambda$ & 150.54 \\
$\delta$ & 1.6369 \\
A & 348.145%
\end{tabular}
\label{Table 1}
\end{table}%
Using these parameters, we notice that $N_{(1)}^{\pm }$ assume complex
values, from what we conclude that the predicted equilibria at $%
(N_{(1)}^{+},0)$ and\ $(N_{(1)}^{-},0)$ are meaningless in the present
context.

The numerical value for $\beta $ is specified separately considering the two
experiments. For the single population case \cite{Dai and Gore Science 2012}%
, we found $\beta =5.831\times 10^{-6}$. This has been obtained\emph{\ }using%
\emph{\ }the parameters in Table 1 and adjusting the two monomorphic
equilibrium $\{(N_{(2)}^{+},1),$\ $(N_{(2)}^{-},1)\}$ at the respective
experimental points $\{(1.75\times 10^{5},1),$\ $(1.5\times 10^{3},1)\}$.
For the experiments performed with both cooperators and cheaters \cite%
{Sanchez and Gore PLOs 2013}, the two-population case, we found \textbf{\ }$%
\beta =1.7185\times 10^{-5}$. These results are in consonance with the
analysis performed by the authors of the experiments: their data indicate
that the carrying capacity of the system with isolated population of
cooperators is larger than the carrying capacity for the mixture of
cooperators and cheaters. In turn, this suggests that under these
experimental conditions, the cooperators are more efficient in using the
available resources in the absence of cheaters.

\subsection{The single population case}

We shall now examine the data from the experiments reported in Ref. \cite%
{Dai and Gore Science 2012}. In these experiments one observes the
properties of \textit{yeast} populations of cooperators, in the absence of \
the cheaters, with respect to changes in the considered \textit{dilution
factor} $\theta .\ $\ From the perspective of the present model, this
corresponds to fix $p=1$ in the above equations and then examine the
behavior of the solutions under changes on $\nu .$

Notice first that for $p=1$ (absence of cheaters) one has $\dfrac{\partial
\overset{\cdot }{p}}{\partial N}=0.$ Thus, the corresponding eigenvalues of $%
\left. J\right\vert _{(N_{(i)},p_{(i)})},$ $i=1,2$ are real. Each of the two
equilibrium points at $p=1,$ i.e. $(N_{(2)}^{\pm },1),$ represents a
monomorphic equilibrium such that%
\begin{equation}
\begin{array}{l}
\left( \tfrac{\partial \overset{\cdot }{N}}{\partial N}\right)
_{(N_{(2)}^{+},1)}=-\Delta _{2}N_{(2)}^{+}<0 \\
\left( \tfrac{\partial \overset{\cdot }{N}}{\partial N}\right)
_{(N_{(2)}^{-},1)}=+\Delta _{2}N_{(2)}^{-}>0%
\end{array}
\label{monomorphic equilibrium}
\end{equation}%
with $\Delta _{2}\equiv \sqrt{\left( \tfrac{1+\left( B-C\right) /\lambda }{A}%
-\beta \right) ^{2}-4\tfrac{\beta \nu }{A\lambda }}$ . From this we conclude
that $(N_{(2)}^{+},1)$ is stable (attracting) whereas $(N_{(2)}^{-},1)$ is
unstable (repelling) equilibrium point.

We then examine the behavior of $N_{(2)}^{+}$ and $N_{(2)}^{-}$, Eqs. (\ref%
{pontos fixos} ($b$)) as functions of the dilution $\nu $, the remaining
parameters fixed as in Table 1. The results are shown in\textbf{\ Figure 1}.
The shape of the curves reproduces the referred experimental study in which
one identifies a \textit{turning point bifurcation} \cite{dynamical systems}%
. Our results predict this bifurcation and explain its origin as occurring
at a value of $\nu $ for which $\Delta _{2}=0$ so that the points $%
(N_{(2)}^{+},1)$ and $(N_{(2)}^{-},1)$ coalesce.

\begin{figure}[ht]
\center
\includegraphics[scale=0.5]{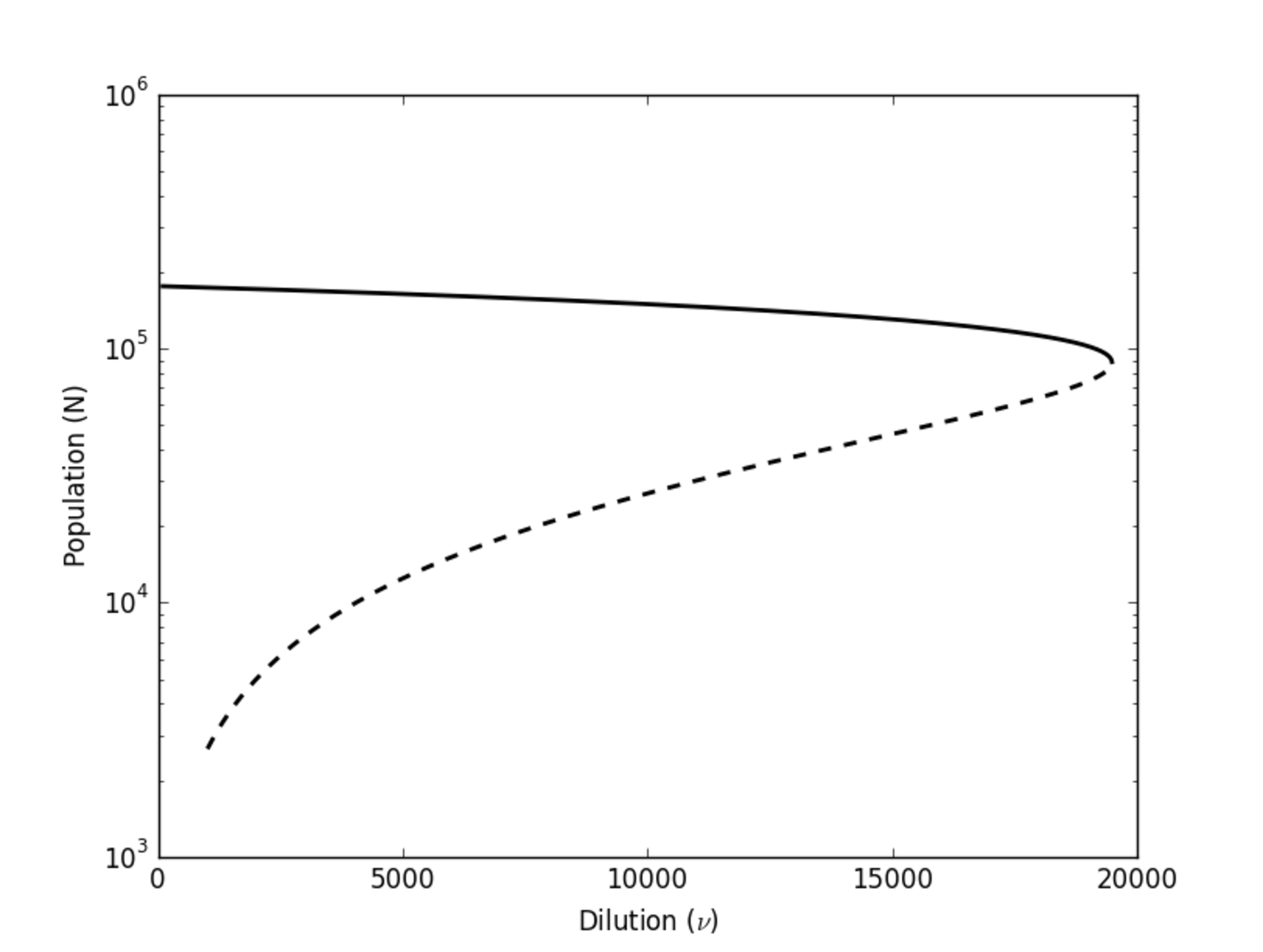}
\caption{The behavior of the equilibrium points $%
(N_{(2)}^{+},1)$ (solid) and $(N_{(2)}^{-},1)$ (dashed) as the parameter $%
\nu $ changes.}
\end{figure}

Let us now examine the time evolution of the density of cooperators $n_{1}(t)
$ for $p=1,$ as solutions to
\begin{equation}
\overset{\cdot }{n_{1}}=n_{1}\left\{ \left( \frac{n_{1}}{A}-1\right)
[(B-C)+(1-\beta n_{1})\lambda ]-\nu \right\} .  \label{dynamics coop 1D}
\end{equation}%
Numerical results for $n_{1}(t),$ obtained from this equation using $\nu
=507.21$ and $\nu =2480.0$ (corresponding to the experimental $\theta =750$
and $\theta =1400,$ respectively) are shown in \textbf{Figures. 2(a)} and
\textbf{2(b)} for several initial population densities $n_{1}(0).$ In
\textbf{Figure 2(a) }we see that the trajectories initiating at relative
high densities approach the stable equilibrium point $(N_{(2)}^{+},1)$,
whereas the populations initiating at densities below the unstable
equilibrium point $(N_{(2)}^{-},1)$ go extinct. Because $N_{(2)}^{-}$
increases with $\nu $ (see expression (\ref{pontos fixos}) ($b$)) we explain
in this way the data indicating that increasing $\nu $ may drive population
to extinction even for those high initial values for which it would have
been survived if $\nu $ were lower, see \textbf{Figure 2(b)}.%

\begin{figure}[ht]
\center
\subfigure[$\nu =507.21$]{
    \includegraphics[scale=0.37]{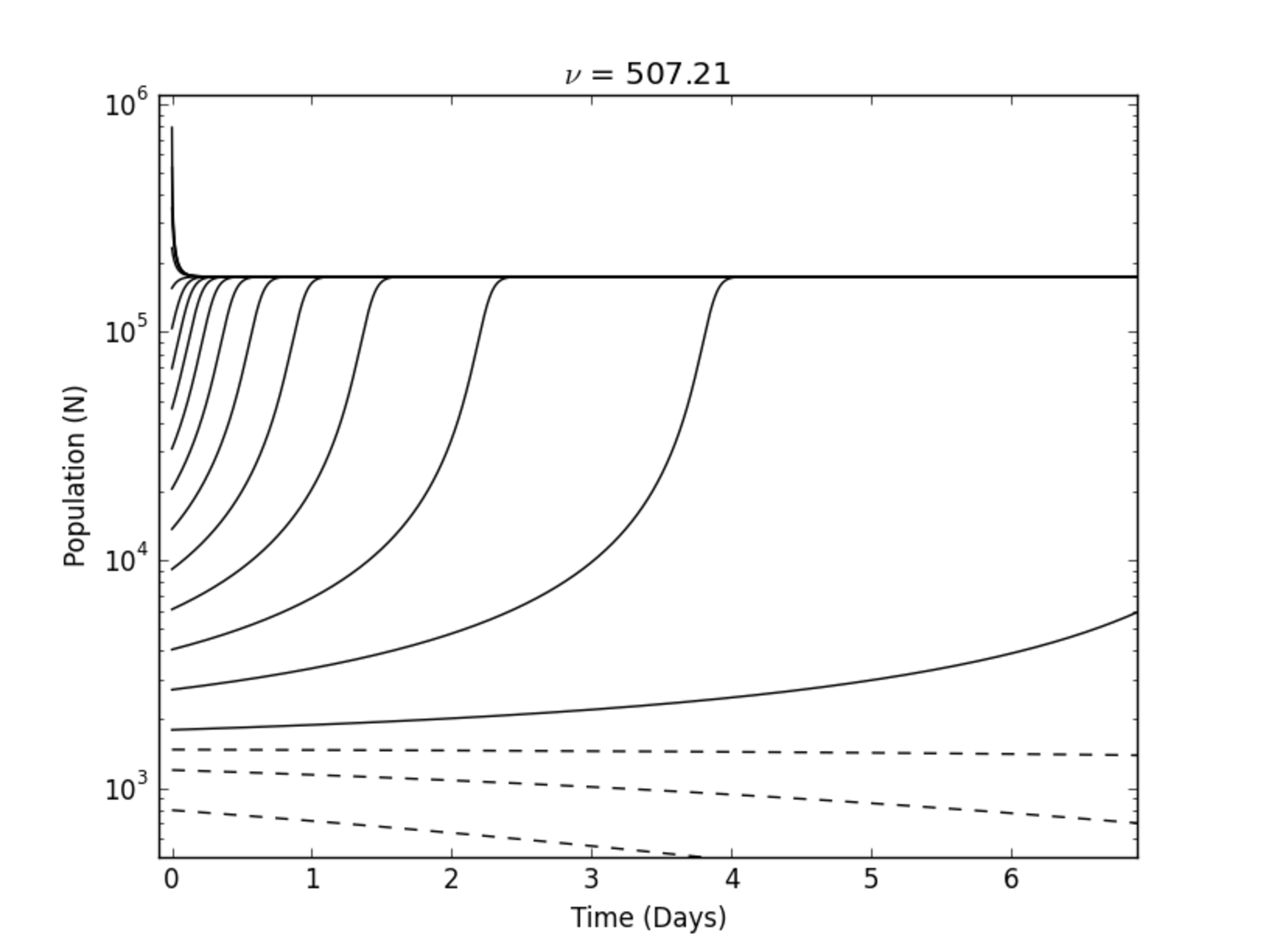}
  } \qquad
\subfigure[$\nu =2480.0.$]{
    \includegraphics[scale=0.37]{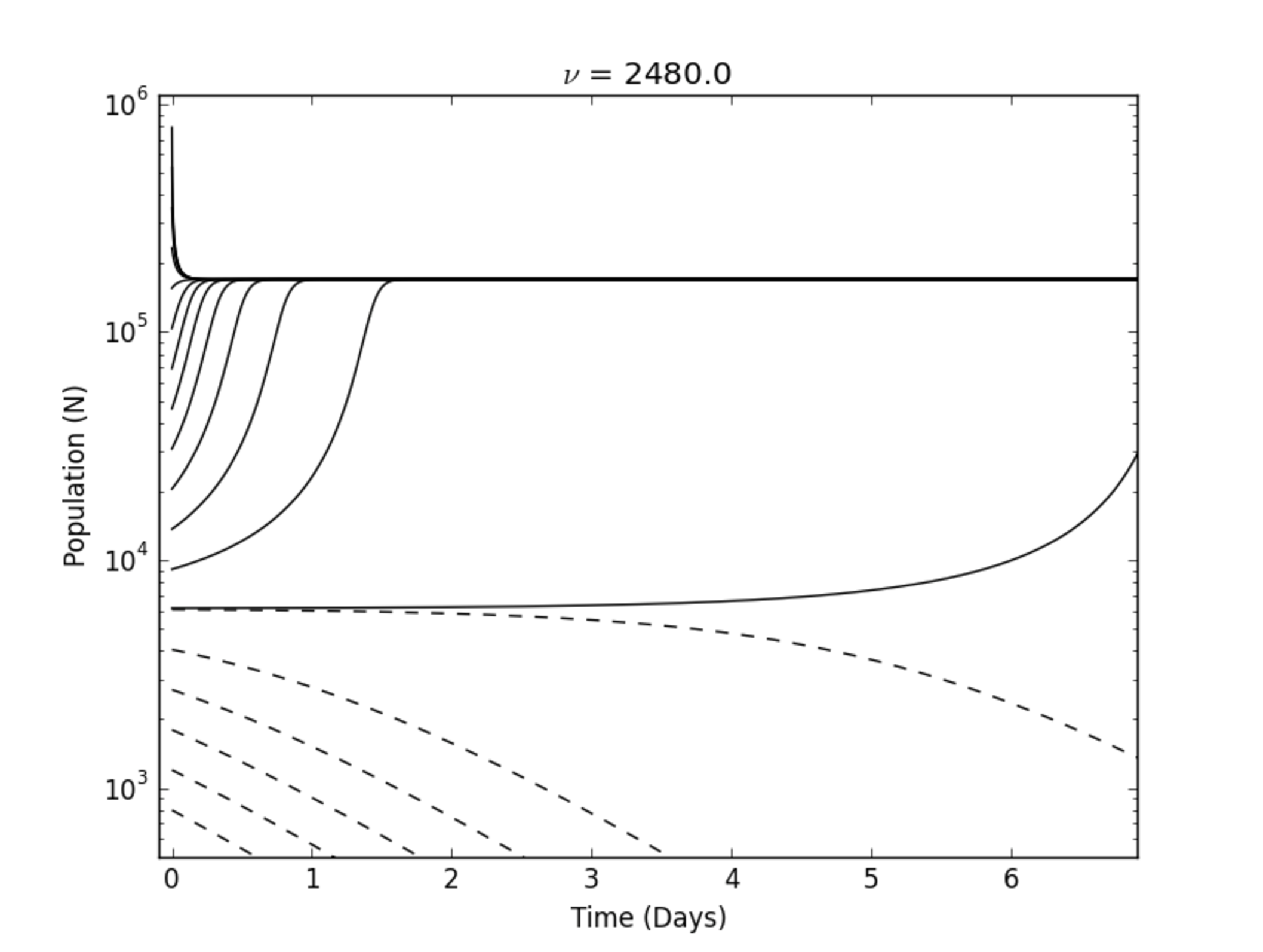}
  }
\caption{Time evolution $n_{1}(t)$ of single populations
of \textit{yeast }cooperators according to Eqs. (\ref{modelo 6}) for $p=1.$
Population go extinct if initial conditions $n_{1}(0)$ are chosen below $%
N_{(2)}^{-}$ (dashed); otherwise population is driven to the stable
equilibrium $N_{(2)}^{+}$ (solid)}
\end{figure}

The results in \textbf{Figures 1 }and\textbf{\ 2 }suggest that the model
proposed here describes very accurately the behavior of the single
populations of cooperator \textit{yeast}, as studied in \cite{Dai and Gore
Science 2012}. In turn, this supports the idea that the analytical results
for $N_{(2)}^{+}$ and $N_{(2)}^{-}$ as functions of the parameters may be
useful to foresee the conditions leading to the critical slowing down of
these populations under external perturbations, as expressed in the
experiments through the dilution protocol.

\subsection{The two population case}

We use the numerical values in Table 1, also $\beta =1.7185\times 10^{-5}$
as specified above, and $\nu =73.9$ to proceed into the phenomenological
analysis of the experimental data in Ref. \cite{Sanchez and Gore PLOs 2013}.
A global view of the $N\times p$ phase space of the model, Eqs. (\ref{dN/dt}%
) and (\ref{dp/dt}), can be achieved by means of the \textit{nullclines}
represented in \textbf{Figure 3. }The $N$\textit{-nullclines} are obtained
from the solutions to the equation (\ref{dN/dt}) for\ $\overset{\cdot }{N}=0,
$ which are: i) the line $N=0$ (not represented) and ii) the curve defined
by the set of points $(N,p)$ satisfying%
\begin{equation}
\left( \frac{N}{A}-1\right) [p(B-C)+(1-\beta N)(p(\lambda -\delta )+\delta
)]-\nu =0.  \label{N_nullcline}
\end{equation}%
A trace of this curve is shown in \textbf{Figure 3} (solid). The $p$\textit{%
-nullclines} also depicted on the same figure (dashed) correspond to the
solutions to the equation \ $\overset{\cdot }{p}=0$, resulting $p=0$ , $p=1$%
, $N=A$ and $N=(\lambda -\delta -C)/(\beta (\lambda -\delta ))$.

\begin{figure}[ht]
\center
\includegraphics[scale=0.5]{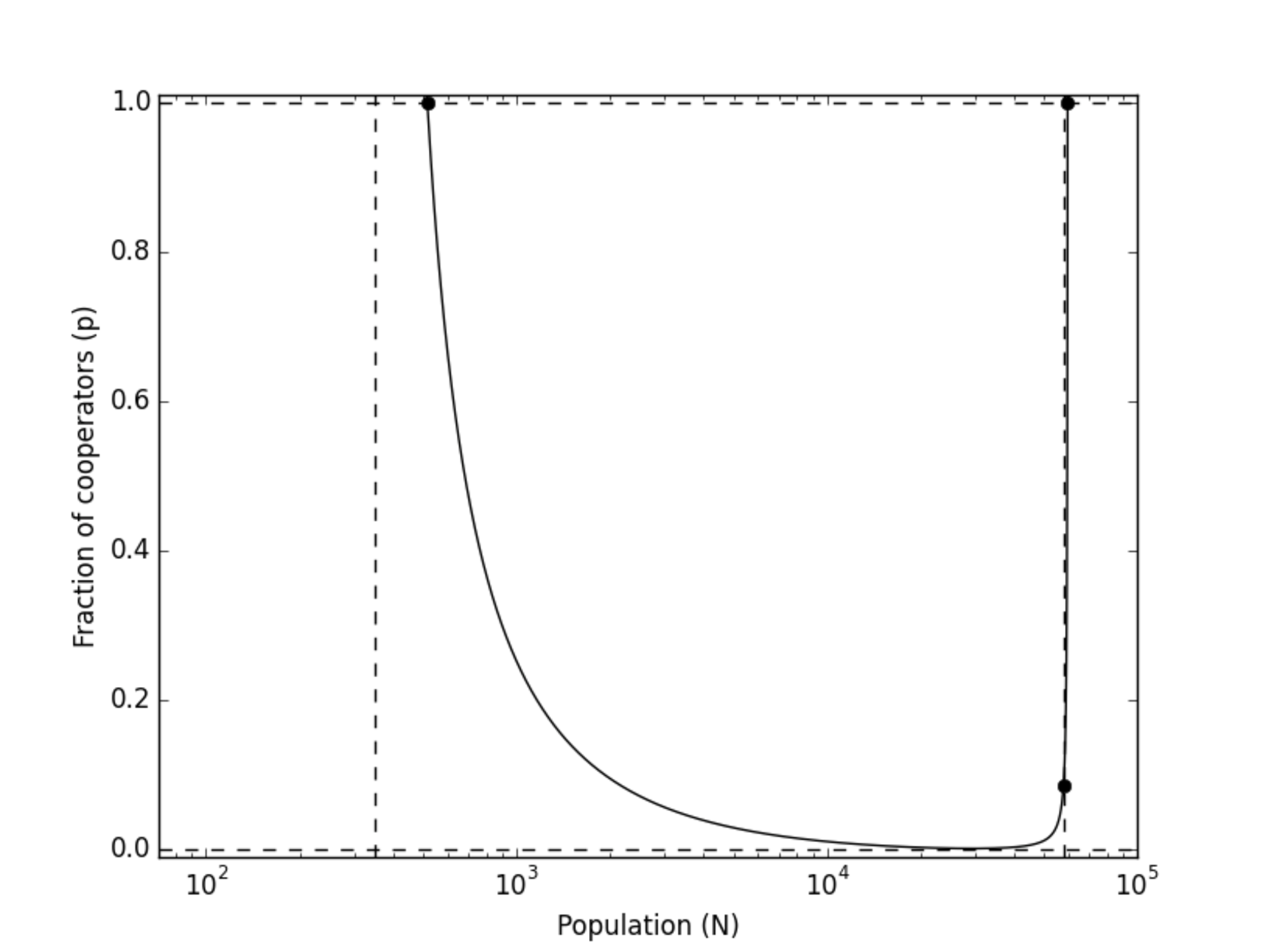}
\caption{The set of $N$-nullclines (solid) and the $p$%
-nullclines (dashed) represented on the $(N\times p)$ phase diagram.}
\end{figure}

The entire set of equilibrium points predicted by the model are depicted at
the encounters of the $p$-nullclines with the $N$-nullclines. It is worth
noticing in these results that the shape of curve defined by the set of
points satisfying (\ref{N_nullcline}), seems to approach very well the trace
of the curve identified by the authors of the experiments as a \textit{%
separatrix} \cite{Sanchez and Gore PLOs 2013}. In fact, our results indicate
that starting from values above this curve and for $p<1$ the population is
driven to the internal equilibrium at $(N_{(3)},p_{(3)})$ representing
\textit{coexistence} of cheaters and cooperators. This is illustrated in
\textbf{Figure 4,} which shows the outcomes of numerical solutions to (\ref%
{dN/dt}) and (\ref{dp/dt}) for the time evolution $N(t)$ and $p(t)$,
represented parametrically on the phase diagram $N\times $ $p$ for several
initial conditions $(N(0),p(0)).$

\begin{figure}[ht]
\center
\includegraphics[scale=0.5]{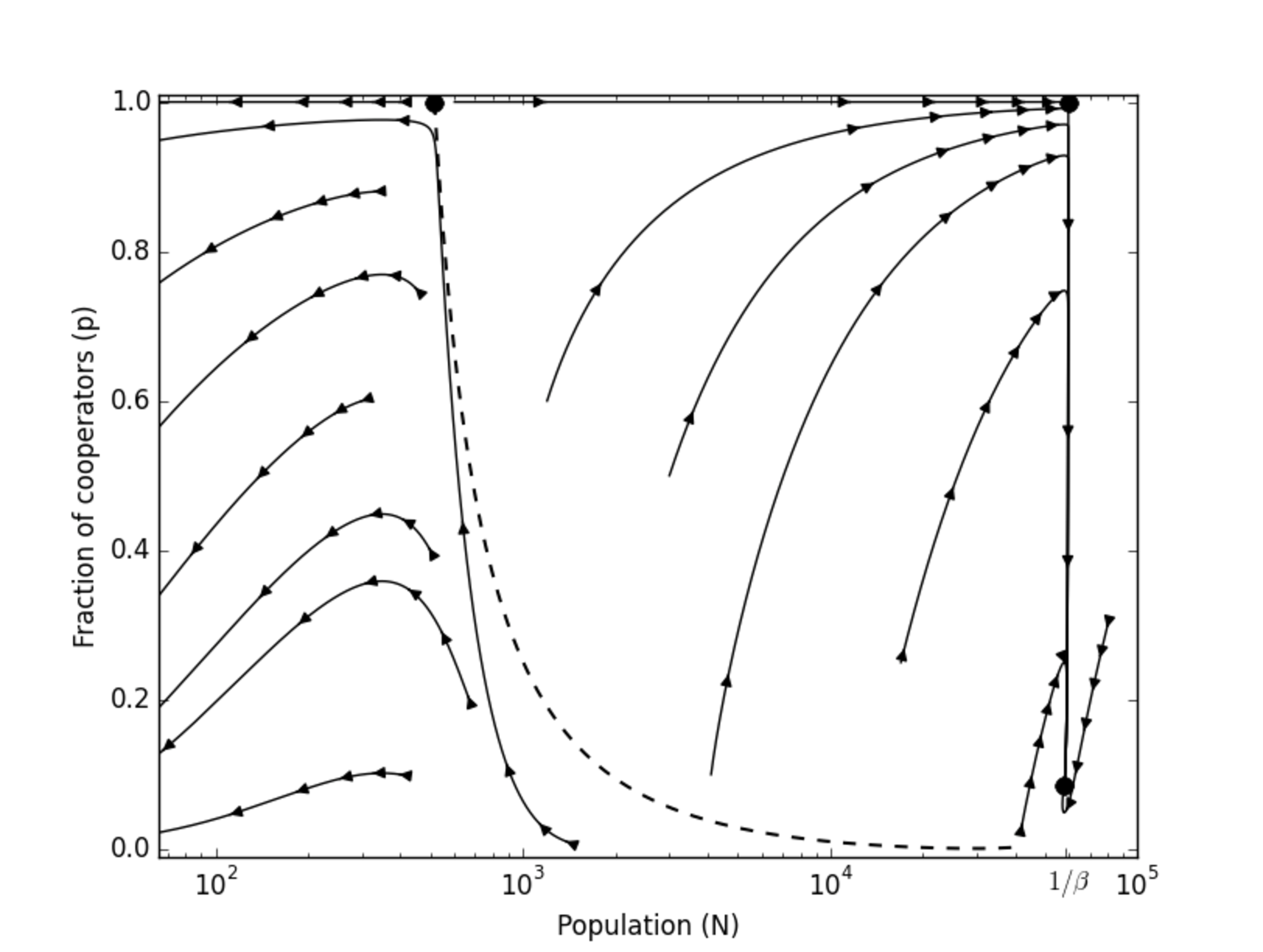}
\caption{Time evolution of mixed populations of \textit{yeast%
} represented parametrically on the $(N\times p)$ phase space for diverse
initial conditions $(N(0),p(0))$. The arrows point to the direction of the
dynamics as the time increases. The two points used as references are $N_{3}=57800$ and $\beta ^{-1}=58190$ (indicated).}
\end{figure}

The complete phase diagram of the model is represented in \textbf{Figure 5}.
These results should be compared qualitatively and also quantitatively with
the corresponding diagrams obtained experimentally for the two population
case.

\begin{figure}[ht]
\center
\includegraphics[scale=0.4]{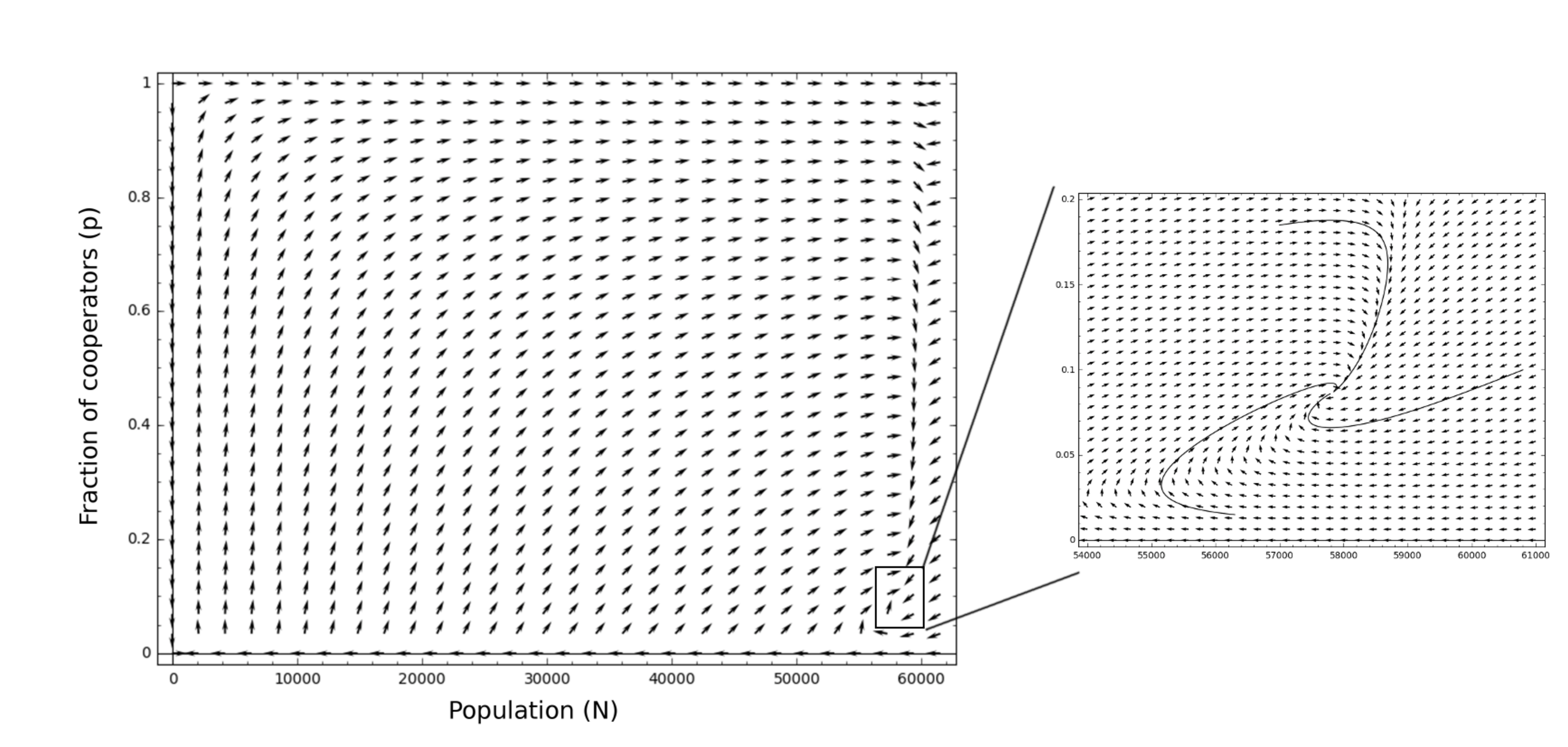}
\caption{The complete phase diagram of the model. The
insert shows the field near the internal equilibrium point represented by%
\emph{\ }the simulated trajectories that illustrate the spiraling dynamics
following different initial conditions.}
\end{figure}

The nature of the internal equilibrium point $(N_{(3)},p_{(3)})$ shall be
disclosed through the eigenvalues $\Gamma ^{\pm }$ of the Jacobian matrix $%
\left. J\right\vert _{(N_{(3)},p_{(3)})}$ (\ref{jacobiano np}) evaluated at
the referred point. Defining $Q_{(3)}=\left( \frac{N_{(3)}}{A}-1\right)$,
we write%
\begin{equation}
\left. J\right\vert _{(N_{(3)},p_{(3)})}=
\begin{pmatrix}
\dfrac{N_{(3)}}{Q_{(3)}}\left[\dfrac{\nu }{A}-\beta (\lambda p_{(3)}+\delta
(1-p_{(3)}))(Q_{(3)})^{2}\right] & N_{(3)}Q_{(3)}B \\
-\beta p_{(3)}(1-p_{(3)})Q_{(3)}(\lambda -\delta ) & 0%
\end{pmatrix}%
\label{jacobiano 3}
\end{equation}%
and the corresponding eigenvalues as $\Gamma ^{\pm }=X\pm \frac{1}{2}\sqrt{%
\Delta },$ where%

\begin{equation}
\begin{array}{l}
X=\dfrac{N_{(3)}}{2Q_{(3)}}\left[\dfrac{\nu }{A}-\beta (\lambda p_{(3)}+\delta
(1-p_{(3)}))(Q_{(3)})^{2}\right]\qquad \\
\\
\Delta =\left( \dfrac{N_{(3)}}{Q_{(3)}}\left[\dfrac{\nu }{A}-\beta (\lambda
p_{(3)}+\delta (1-p_{(3)}))(Q_{(3)})^{2}\right]\right)
^{2}-4N_{(3)}p_{(3)}(1-p_{(3)})B\beta (\lambda -\delta )(Q_{(3)})^{2}%
\end{array}
\label{eigenvalue parts X and delta}
\end{equation}

Since $p=p_{(3)}$ is the only non-trivial \textit{equilibrium} solution for
the frequency of cooperators (i.e., different from $p=0$ or $p=1$), it is
suggestive to look for the conditions under which $(N_{(3)},p_{(3)})$
exhibit properties of an \textit{attractive node }as identified in the
experimental data, and also as suggested in \textbf{Figure 5} (see \textit{%
insert}). This requires $\Delta <0$ and $X<0$. Indeed, these two conditions
are fulfilled by the numerical choices for the parameters as in \textbf{%
Table 1}.

In retrospective, this means that by identifying the coordinates of the
equilibrium points predicted by the model with the corresponding
experimental, one concludes that the internal equilibrium point predicted
theoretically is an attractive node, precisely as observed. We can also
obtain from our results that in contrast to the equilibrium points at $p=1$
there is no bifurcation associated with $(N_{(3)},p_{(3)})$ since it remains
attractive for all positive $\nu $.

\section{Discussion}

Observe that for equal growth rates $\lambda =\delta ,$ Eqs. (\ref{dN/dt})
and (\ref{dp/dt}) present no solutions for internal equilibrium points$.$
The invasion condition for fixation of cooperation, namely $\frac{1}{n_{1}}%
\frac{dn_{1}}{dt}>\frac{1}{n_{2}}\frac{dn_{2}}{dt}$ is equivalent to $%
(\lambda -\delta )(1-\beta N)>C$ which can be satisfied if $\lambda \neq
\delta $ for a certain range of parameters in all regions of the phase plane
for which $N<1/\beta .$ Conversely, even taking $\lambda \neq \delta $, and
also $B\neq C$ such that $B,C>0,$ but in the absence of the factor
controlling Allee effects, one would still find a solution to the internal
equilibrium $(N_{(3)},p_{(3)})$ although in this case the bifurcation
predicted at the projection line $p=1$ would be suppressed.

The existence of the attractive node as a result from an eco-evolutionary
feedback can be understood by examining the behavior of the different
contributions to Eqs. (\ref{dN/dt}) and (\ref{dp/dt}). For $N\gtrsim 1/\beta
$, the environment effects weighted by the difference between the intrinsic
growth rates $(\lambda -\delta )(1-\beta N)$ dominates the dynamics: the
entire population and in particular the density of cooperators in the
population decrease fast since $(\lambda -\delta )$ is large and positive.
As $N$ decreases reaching values such that $N\lesssim 1/\beta ,$\emph{\ }then%
\emph{\ }$N$ continues decreasing while dilution effects dominates. The
frequency $p$ however, can revert its behavior since the cost $-C$
represented in (\ref{dp/dt}) contributing negatively to the fitness of
cooperators can be overcomed by the growth rate contribution. As $p$
increases, both the game and intrinsic growth rate contributions in Eq. (\ref%
{dN/dt}) will eventually overcome the effects of dilution so that the
population as a whole increases until reaching the carrying capacity. At
this point the behavior of $N$ would follow the sum of the contribution due
to the game, namely $p(B-C)$ and the one due to dilution. While this
combination assumes positive values, $N$ continues to increase until $%
(\lambda -\delta )(1-\beta N)$ dominates the dynamics with a large negative
contribution and the feedback cycle resumes. What is important to notice is
that this behavior is largely ancored on the process of dilution and on the
diference between the intrinsic growth rates. This comes in addition to the
effects more often discussed in the literature regarding the contributions
to the fitness due to the Lotka-Volterra and game at different regions of
the phase space.

The behavior of the solutions near $(N_{3},p_{3})$ can also be investigated
from the point of view of the two populations, equations (\ref{modelo 6}).
For this we define an \textit{effective game, }with payoffs $\overline{a_{ij}%
}$ $i,j=1,2$ being functions of both $N$ and $\nu ,$ such that%

\begin{equation}
  \bordermatrix{~ & ECo & ECh \cr
                    ECo & \overline{a_{11}}=\left[ B-C+\lambda (1-\beta N)\right]
                    (\tfrac{N}{A}-1)-\nu \text{ \ \textbf{\ }} & \overline{a_{12}}=\left[
                    -C+\lambda (1-\beta N)\right] (\tfrac{N}{A}-1)-\nu \cr
                    ECh & \overline{a_{21}}=\left[ B+\delta (1-\beta N)\right] (%
                    \tfrac{N}{A}-1)-\nu & \overline{a_{22}}=\left[ \delta
                    (1-\beta N)\right] (\tfrac{N}{A}-1)-\nu \cr}
\label{N and ni dependent payoff}
\end{equation}

where \textit{ECo} and \textit{ECh }refer to effective cooperation and
effective defect (cheat) strategies, respectively. With these definitions we
rewrite Eqs. (\ref{modelo 6}) as

\begin{equation}
\begin{array}{l}
\overset{\cdot }{n}_{1}=n_{1}\left[ \overline{a_{11}}p+\overline{a_{12}}(1-p)%
\right] \equiv n_{1}F_{1}(N,p) \\
\\
\overset{\cdot }{n}_{2}=n_{2}\left[ \overline{a_{21}}p+\overline{a_{22}}(1-p)%
\right] \equiv n_{2}F_{2}(N,p)%
\end{array}
\label{modelo 6 efetivo com diluicao}
\end{equation}%
expressing in this way the fitness for cooperators $F_{1}(N,p)$ and for
cheaters $F_{2}(N,p)$ in terms of the effective payoffs $\overline{a_{ij}}$.
In turn, the time variations $\overset{\cdot }{N}$ and $\overset{\cdot }{p}$
can be expressed in terms of $F_{1}$ and $F_{2}$ as
\begin{equation}
\begin{array}{l}
\overset{\cdot }{N}=n_{1}F_{1}+n_{2}F_{2} \\
\\
\overset{\cdot }{p}=p(1-p)(F_{1}-F_{2})%
\end{array}
\label{variacoes p e N em termos de F1 e F2}
\end{equation}%
Because we want to examine the behavior of this effective game near $%
(N_{(3)},p_{(3)})$ we observe in particular that%
\begin{equation}
\begin{array}{l}
\overline{a_{11}}(N_{(3)})=\overline{a_{21}}(N_{(3)})=\left( B+C\dfrac{%
\delta }{\lambda +\delta }\right) Q_{(3)}-\nu \\
\\
\overline{a_{12}}(N_{(3)})=\overline{a_{22}}(N_{(3)})=\left( C\dfrac{\delta
}{\lambda +\delta }\right) Q_{(3)}-\nu .%
\end{array}%
\text{and}  \label{payoffs iguais}
\end{equation}%
From this, it follows that $F_{1}(N_{(3)},p)=F_{2}(N_{(3)},p)$ for all $p\in %
\left[ 0,1\right] $. Evidently, for variable $N$ \ the condition $%
F_{1}=F_{2} $ is not sufficient to characterize equilibrium as it would be
the case of the \textit{replicator dynamics} for which $N$ is a conserved
quantity \cite{Nowak livro}. For variable $N$, the required condition for
equilibrium is that $F_{1}=F_{2}=0$ which is fulfilled at the point $%
(N_{(3)},p_{(3)})$.

The four entries of the effective payoff matrix (\ref{N and ni dependent
payoff}) are represented in \textbf{Figure 6} as functions of $N,$ near $%
N_{(3)}.$ The parameters used in the plot are those from in Table 1, $\beta
=1.7185\times 10^{-5}$ and $\nu =73.9$. Observe that the line $N=N_{(3)}$
(dashed) sets out two regions within each of which the effective payoffs
characterize different games. For $N\gtrsim N_{3}$ the inequalities $%
\overline{a_{21}}(N)>\overline{a_{11}}(N)>\overline{a_{22}}(N)>\overline{%
a_{12}}(N)$ hold so that in this region the effective game corresponds to a
PDG for which the \textit{effective defect} is an ESS (Evolutionary Stable
Strategy), the dominant strategy for both players. In the region for which $%
N\lesssim N_{(3)}$ one has $\overline{a_{11}}(N)>\overline{a_{21}}(N)>%
\overline{a_{12}}(N)>\overline{a_{22}}(N)$ corresponding to a \textit{%
Harmony game} for which the dominant strategy for both players is the
\textit{effective cooperation}, the ESS in this case. At $N=N_{(3)}$ , one
has $\overline{a_{11}}(N_{(3)})=\overline{a_{21}}(N_{(3)})$ and $\overline{%
a_{12}}(N_{(3)})=\overline{a_{22}}(N_{(3)})$ indicating that both effective
defect and effective cooperation are Nash equilibria at the coexistence
condition but none of these is an ESS.

\begin{figure}[ht]
\center
\includegraphics[scale=0.5]{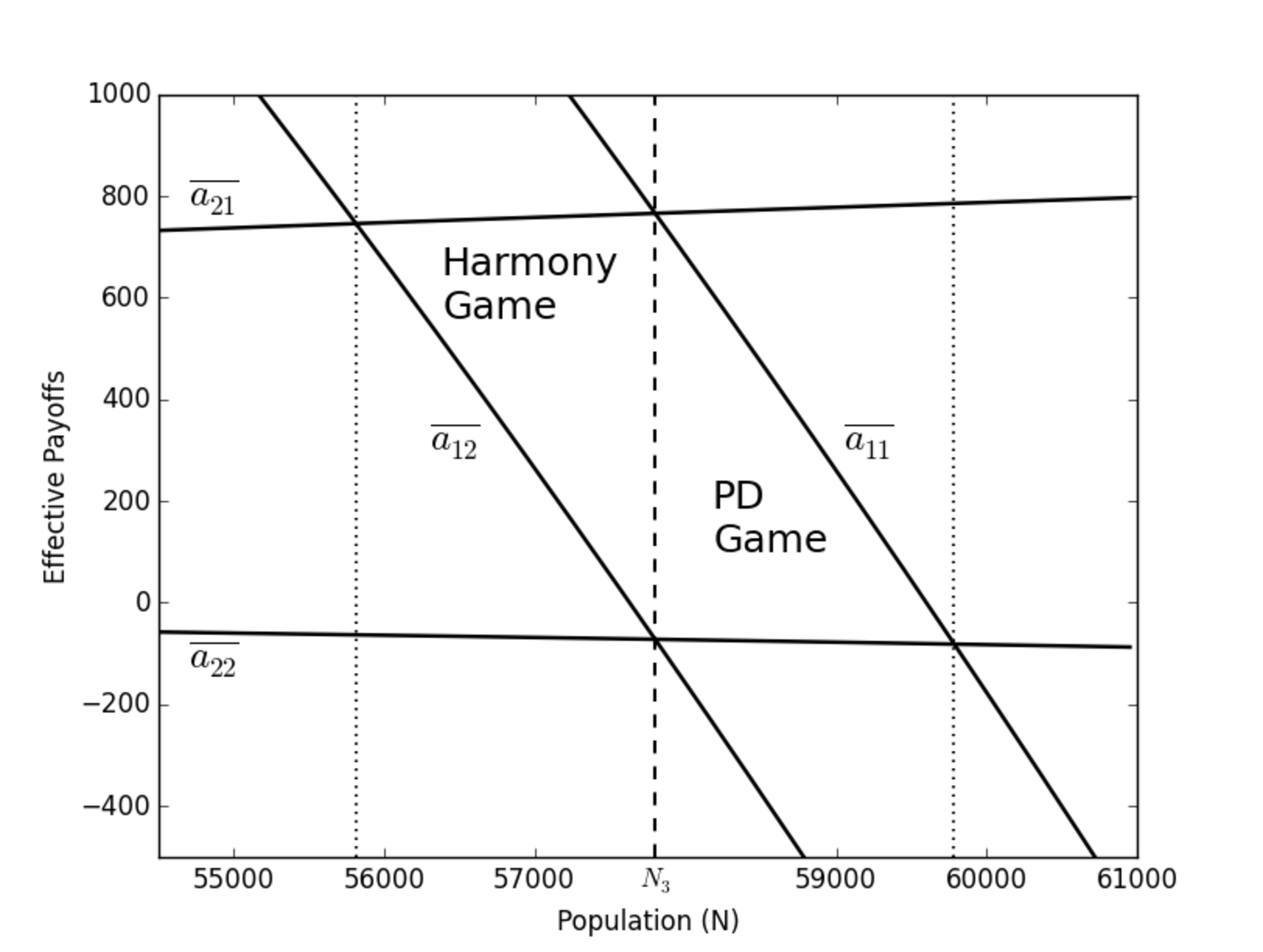}
\caption{ Effective payoffs $\overline{a}_{ij}$ $i,j=1,2$ as
functions of $N$, near $N_{3}$ (dashed). The regions where the strategies
characterize different games are explicitly indicated. For Harmony game it
is limited by $N_{3}$ and the solution $N$ to $\overline{a}_{12}=\overline{a}%
_{21}$ (left dotted). \ For PD Game it is limited by $N_{3}$ and the
solution $N$ to $\overline{a}_{11}=\overline{a}_{22}$ (right dotted).}
\end{figure}

A similar analysis has been performed in Ref. \cite{Requejo Camacho PRL 2012}%
. Although the eco-evolutionary model there has also been conceived in the
absence of structured populations, the correspondence to our analysis is not
immediate because their dynamic payoff matrix has been defined \textit{a
priori}, directly in terms of the resources exchanged between pairs of
individuals. The effective matrix built up here comprises the original game,
the dilution factor and the ecological effects, what allowed us to evaluate
separately each contribution to the dynamics discussed here.

\section{Conclusions}

The phenomenological analysis presented above suggests that the dilution
process introduced into the considered experiments primarily to examine
properties of resilience of the populations under changing environments, may
be crucial for observing coexistence of cooperators and cheaters in the
mixed population of \textit{yeast}. The analytical results have been
achieved in terms of parameters strictly related to the populations traits
or to their interactions with the environment. The strength of dilution is a
control parameter. We believe that this study can be useful to make further
predictions on the properties of such systems under different conditions.
For example, it shall be of interest to examine the limits for population
extinction by analyzing the behavior of the N-nullcline Eq.(\ref{N_nullcline}%
), with respect to changes in the diverse parameters. This curve appears to
approach very well the trace of the separatrix identified in the data.

The analysis of the defined effective game offers an alternative to
understand the outcomes of the dynamics in terms of the nature of strategies
that change within each region of the considered phase space.

\section{Acknowledgements}

The financial support from Funda\c{c}\~{a}o de Amparo \`{a} Pesquisa do
Estado de S\~{a}o Paulo (FAPESP), Brazil under Grant No. 2015/17395-3 is
gratefully acknowledged.

\end{document}